\documentclass[
reprint,
superscriptaddress,
amsmath,
amssymb,
aps,
prb,
notitlepage,
floatfix
]{revtex4-2}

\usepackage{caption}
\usepackage{graphicx}
\usepackage{dcolumn}
\usepackage{bm}
\usepackage{float}
\usepackage{upgreek}
\usepackage{mathptmx}
\usepackage{multirow}
\usepackage[sfdefault,lf]{carlito} % Sans Serif font
\usepackage{setspace} % For line spacing
\begin{document}

\title{Electron Tesla valve}

\author{Daniil I. Sarypov}
\email{d.sarypov@g.nsu.ru}
\author{Dmitriy A. Pokhabov}
\author{Arthur G. Pogosov}
\author{Evgeny Yu. Zhdanov}
\affiliation{Rzhanov Institute of Semiconductor Physics SB RAS, Novosibirsk, 630090, Russia}
\affiliation{Novosibirsk State University, Novosibirsk, 630090, Russia}
\author{Andrey A. Shevyrin}
\affiliation{Rzhanov Institute of Semiconductor Physics SB RAS, Novosibirsk, 630090, Russia}
\author{Askhat K. Bakarov}
\affiliation{Rzhanov Institute of Semiconductor Physics SB RAS, Novosibirsk, 630090, Russia}
\affiliation{Novosibirsk State University, Novosibirsk, 630090, Russia}

\begin{abstract}
In solids, frequent electron-electron collisions can induse collective, fluid-like electron transport. While this regime offers a powerful framework for exploring many-body phenomena, there is still a lack in functional electronic device actively exploting hydrodynamic behaviour of electrons. Here, we introduce a solid-state analogue of a Tesla valve --- a passive fluidic diode that rectifies flow without moving parts. Lithographically defined in high-mobility GaAs two-dimensional electron gas, the device exhibits abrupt rectification producing a more than tenfold difference between forward and reverse resistances. This threshold behaviour, reminiscent of the onset of turbulence in fluidic Tesla valves, points to the emergence of turbulent regime in the electron liquid --- a long-predicted, but yet unobserved state of electronic matter. More broadly, our work demonstrates the fruitfulness of the hydrodynamic analogy: fluidic technologies can be readily adopted to create novel electronic devices. Here, this is realized through a solid-state rectifier whose operation relies on a new physical mechanism, interparticle collisions.
\end{abstract}

\maketitle
\normalsize

In conventional theories of electron transport in solids, electrical resistance is governed by momentum-relaxing scattering processes: interactions with impurities, defects and phonons. Electron-electron (\textit{e-e}) collisions, although frequent, are typically considered secondary because they conserve total momentum of the electron system and therefore do not directly contribute to resistivity. In sufficiently clean conductors, however, \textit{e-e} collisions can dominate the charge carrier dynamics, and electron system behaves as a collective medium whose motion can be described within a hydrodynamic framework \cite{gurzhi1968, fritz2024}. Recent experiments have confirmed this picture, revealing many-body phenomena characteristic of viscous fluids, including Poiseuille electron flow in microchannels \cite{sulpizio2019, ku2020, vool2021}, current vortices \cite{bandurin2016, aharon-steinberg2022, palm2024} and superballistic transport \cite{guo2017, krishnakumar2017, ginzburg2021, kravtsov2024, estrada-álvarez2025b, sarypov2025}.

The hydrodynamic analogy between electron liquids and classical fluids has thus proven remarkably fruitfull, offering intuitive insight into collective phenomena and suggesting new routes for engineering interaction-dominated transport. Yet an important question remains: how far can this analogy be extended? In particular, can it encompass highly nonlinear phenomena such as turbulence, one of the central problems of fluid dynamics?

\begin{figure*}[ht!]
    \centering
    \includegraphics{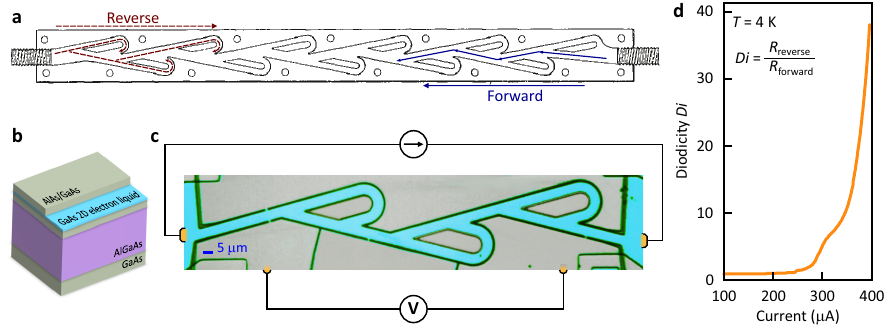}
    \caption{\textbf{GaAs Tesla valve for electron liquid.} \textbf{a,} Schematics of original design adopted from N. Tesla's patent \cite{tesla1920}. \textbf{b,} GaAs/AlGaAs heterostructure with two-dimensional electron liquid in a GaAs quantum well. \textbf{c,} False-color optical micrograph of one of the created valves, overlaid with the four-probe measurement schematic. Scale bar is $5$~$\mu$m. \textbf{d,} Diode efficiency $Di$ for device shown in \textbf{b}, measured as the ratio of reverse to forward resistance values.}
    \label{fig:1}
\end{figure*}

A natural way to adress this question is to transpose well-known fluidic devices into electronic systems and test whether they operate similarly. An especially intriguing candidate is Tesla valve --- a fixed-geometry passive valve that allows fluid to flow more easily in one direction than the other, without any moving parts. Its design consists of a series of interconnected, teardrop-shaped loops [Fig.~\ref{fig:1}a]. Patented over a century ago \cite{tesla1920}, it remains actively studied for its complex internal flow physics \cite{porwal2018, raffel2021, nguyen2021} and potential applications in heat transfer \cite{li2023, huang2024a}. The next logical step is to merge hydrodynamic behaviour of electrons in solids with unique properties of Tesla valve by exploring electron transport in structures that mimic its fluidic counterparts. Beyond fundamental interest, an electronic analogue of fluidic Tesla valve could be a promising candidate for THz applications among other geometric diodes \cite{han2025}. However, the concept of an "electron Tesla valve" is almost unexplored: the only study in graphene  reports \cite{geurs2020} a vanishing rectification of $\sim5$~\%, which is intriguing but far form capturing the full depth of the Tesla valve behaviour.

In this work, we introduce the electron Tesla valve created from GaAs [Figs.~\ref{fig:1}b,c] --- a material where the fluid-like behaviour of electrons is widely reported \cite{dejong1995, gusev2018, levin2018, gupta2021, ginzburg2021, keser2021, wang2022, wang2023, sarypov2025b}. Faithful to Tesla's original geometry \cite{tesla1920}, the devices consist of multiple loops and exhibit strong rectification, with reverse resistance exceeding forward resistance by more than an order of magnitude [Fig.~\ref{fig:1}d]. Notably, the rectification features closely mirrors those of turbulent fluidic Tesla valve \cite{nguyen2021}, while achieved diodicity values substantially higher than those reported in classical fluidic devices---pointing to the emergence of a turbulent regime in the electron liquid. These findings underscore the universality of hydrodynamic description: the Tesla valve operation relies on the same physical mechanism whether it is driven by water or an electron liquid. Beyond fundamental insights into non-Ohmic electron dynamics, they open pathways toward functional hydrodynamic electronic components.

\textbf{GaAs Tesla valves.}
The devices are fabricated on the basis of GaAs/AlGaAs heterostructures with high-mobility two dimensional electron gas (2DEG) [Fig.~\ref{fig:1}b, Extended data Fig.~\ref{ext_data:1}a]. At $T=4$~K, 2DEG has a density of $n=6.8\times10^{11}$~cm$^{-2}$, determined from Shubnikov-de Haas oscillations [Extended data Figs.~\ref{ext_data:1}b-d], and mobility $\mu\approx 10^{6}$~cm$^{2}$/(V$\cdot$s), measured in macroscopic samples made of the same heterostructure (see "Methods"). Driving the electron system into the hydrodynamic regime requires frequent \textit{e-e} collisions, which is achieved by raising the temperature of electron system --- either by heating the lattice (increasing the sample temperature) or by passing a DC current \cite{dejong1995}. On the other hand, raising the temperature intensifies electron-phonon scattering, which breaks momentum conservation, i.e. hydrodynamic regime, and restores conventional Ohmic transport. For high-mobility 2DEG in GaAs quantum wells, this leaves a window below approximately $40$~K \cite{gusev2018, levin2018, sarypov2025b} where \textit{e-e} collisions are frequent but electron-phonon scattering is still negligible allowing hydrodynamic effects to be clearly resolved. Building on these conditions, we performed transport measurements in the extended temperature range of $4$---$70$~K and DC current up $\sim10^{-3}$~A.

\begin{figure*}[ht]
    \centering
    \includegraphics{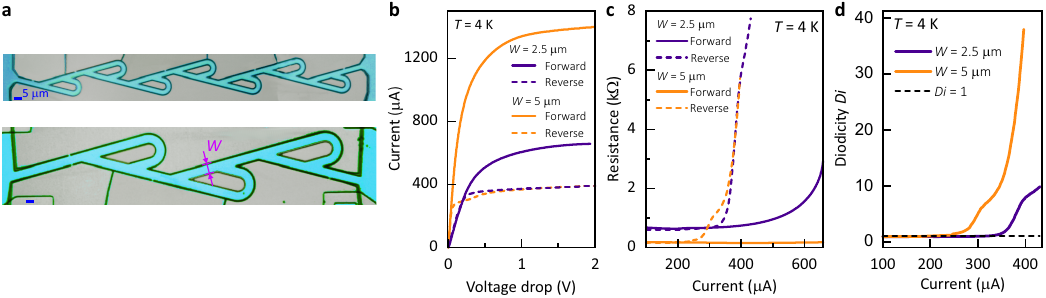}
    \caption{\textbf{Electron transport in Tesla valves.} \textbf{a,} False-color optical micrograph of GaAs Tesla valves of different widths. \textbf{b,} \textit{I-V} characteristics of the devices shown in \textbf{a} at lattice temperature of $T=4$~K. \textbf{c,d} Resistance of Tesla valves and diodicity $Di$ as functions of DC current at lattice temperature of $T=4$~K.}
    \label{fig:2}
\end{figure*}

The Tesla valves are patterned by electron lithography in the plane of a 2DEG, with charecteristic channel width $W$ of $2.5$ and $5$~$\mu$m, where the wider valve is scaled proportionally from narrower one [Fig.~\ref{fig:2}a]. Micron-scaled width of the samples supports hydrodynamic transport while excluding ballistic mechanisms responsible for rectification in so-called "ballistic rectifiers" \cite{song1998, auton2016, auton2017}. Those devices rely on the collimation or reflection of narrow electron beams and therefore require submicron dimensions. These conditions do not apply to the valves studied here.

In fluidic systems, the rectification of Tesla valves turns on at high enough Reynolds numbers $Re>1$ \cite{nguyen2021, nobakht2013, kim2015, gamboa2005}, corresponding to high flow rates. Here, we follow the approach of Ref.~\cite{nguyen2021}, where the rectification is captured by measuring fluid flow rate as function of pressure drop across the device, and measure  current-voltage (\textit{I-V}) characteristics of the valves across a wide range of electric current [Fig.~\ref{fig:2}b]. In forward direction, where electrons bypass the internal loops [Fig.~\ref{fig:1}a], the \textit{I-V} curve in narrow and wide samples resembles that of a straight macroscopic channel: linear at low bias and saturating at high voltages. The saturation current in both valves corresponds to the same current density of $j=I/W\approx260$~A/m, i.e. the same drift velocity of $v_d=j/(ne)\approx2\times10^5$~m/s. This value is close to the maximal drift velocity reported for electrons in GaAs quantum wells \cite{hirakawa1988, balkan1990, mokerov2009}, indicating that the \textit{I-V} flattening in forward flow direction is due to the drift velocity saturation in studied GaAs structures.

Surprisingly, the electron flow in the reverse direction behaves qualitatively differently [Fig.~\ref{fig:2}b]. Below $I\approx300$~$\mu$A, the \textit{I-V} curves coincide with those corresponding to forward bias, but, at higher currents, they sharply diverge, and electron transport becomes direction-dependent. Notably, lower reverse saturation currents along with the overall shape of the \textit{I-V} curves points on current-limiting physical mechanism distinct from the drift velocity saturation.

\textbf{Rectification effect.} The features of direction-dependent transport are clearly resolved in [Fig.~\ref{fig:2}c] that shows the resistance of the valves $R=V/I$ as function of DC current. Namely, as reverse current exceeds the value of $250$---$350$ $\mu$A, we observe sharp increase of the resistance while the forward resistance remains low ($<1$~k$\Omega$) and almost unchanged indicating linear transport [Fig.~\ref{fig:2}a]. To quantify this behaviour and test the performance of the valves, we evaluate the diodicity $Di$, which is a standard characteristics of asymmetric resistors defined as the ratio of reverse to forward resistance values \cite{li2015, nguyen2021}:

$$Di=\frac{R_\mathrm{reverse}}{R_\mathrm{forward}}.$$
Accordingly, the valve rectifies the flow if $Di>1$. In the studied valves, rectification emerges abruptly and the diodicity rising steeply to values as high as $Di\approx40$ in wider valve [Fig.~\ref{fig:2}d]. An important feature of the diodicity curves is a plateau. A similar structure is also observed in fluidic Tesla valves \cite{nguyen2021} where sharp threshold of diodicity followed by the plateau is attributed to the onset of turbulence. The $Di$ plateau can be a signature of transition process. For water flow in a Tesla valve, such a plateau corresponds to the progressive filling of the structure with intermixed turbulent flows as $Re$ number increases \cite{nguyen2021}. Once the valve is fully saturated with turbulence, the transition concludes and diodicity continues to grow. The studied electron liquid appears to exhibit the same behaviour: the diodicity curve in the electron Tesla valve qualitatively reproduces that of its fluidic analogue. This close similarity points to the universal rectification mechanism in both fluidic and electronic Tesla valves --- the turbulence of hydrodynamic flow.

Such a scenario is plausible since the geometry of the Tesla valve itself favours head-on interparticle collisions, which shifts the onset of turbulence to lower Reynolds numbers $Re$ \cite{nguyen2021}. In electron systems, $Re$ can be large enough, especially at a high DC current that drives the system deeper into hydrodynamic regime by raising electron temperature with respect to the lattice \cite{dejong1995}. Our estimates give a Reynolds number $Re\gtrsim7$ at $I\gtrsim350$~$\mu$A (see "Methods" section for details), which is well above unity. Earlier simulations \cite{mendoza2011, coelho2017} predict the onset of electron turbulence at higher $Re\gtrsim20$ in flow past circular obstacle. We note, however, that the critical Reynolds number is highly sensitive to device geometry. In a Tesla valve, with its characteristic looped shape, the emergence of turbulence is expected at substantially lower $Re$ values, which is confirmed in water-flow experiments \cite{nguyen2021}. Our estimates therefore support the interpretation that turbulent dynamics underlies the observed rectification in electron Tesla valves.

\begin{figure}[h!]
    \centering
    \includegraphics{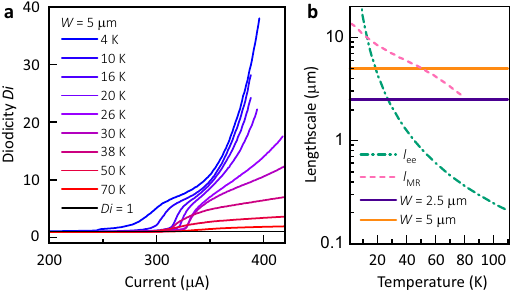}
    \caption{\textbf{Temperature dependence of the diodicity.} \textbf{a,} Diodicity $Di$ of wide ($W=5$~$\mu$m) Tesla valve at different lattice temperatures. \textbf{b,} Characteristic lengthscales of studied system: widths $W$ of the Tesla valves , momentum relaxing length $l_\mathrm{MR}$, \textit{e-e} scattering length $l_{ee}$.}
    \label{fig:3}
\end{figure}

To test the robustness of the rectification, we measured the diodicity at various lattice temperatures [Fig.~\ref{fig:3}a]. 
As the temperature increases from $4$ to $\approx 20$~K, the diodicity plateau smoothly weakens while the curves remain almost converged at $Di>10$. Further, at $T\gtrsim20$~K, the diodicity is significantly suppressed until rectification is almost vanished at $T=70$~K. This suppression can be explained by growing influence of electron-phonon scattering, which disrupts the collective flow and, thus, the rectification effect. The strength of this scattering is quantified by the momentum relaxing length $l_\mathrm{MR}$, which we measure independently in a macroscopic Hall bar made of the same heterostructure as Tesla valves (Fig. \ref{fig:3}b, "Methods"). Below $T<20$~K, $l_\mathrm{MR}$ exceeds $10$~$\mu$m, which is well above the Tesla valve width $W$, meaning that electrons rarely loose their momentum on phonons, and diodicity remains high. Above $T>20$~K, $l_\mathrm{MR}$ becomes comparable to or smaller than $W$, showing increased electron-phonon scattering rate, that leads to expected suppression of diodicity. Notably, \textit{e-e} scattering remains frequent even at low lattice temperature due to high DC current which heats electron system \cite{dejong1995, wang2023}. At $I=350$~$\mu$A, we estimate electron temperature $T_e\approx 100$~K (see "Methods"), where \textit{e-e} scattering length drops below $0.5$~$\mu$m, ten times lower than $W$ [Fig.~\ref{fig:3}b]. This ensures that \textit{e-e} collisions are abundant reinforcing the many-body origin of the rectification.

\begin{figure}[h!]
    \centering
    \includegraphics{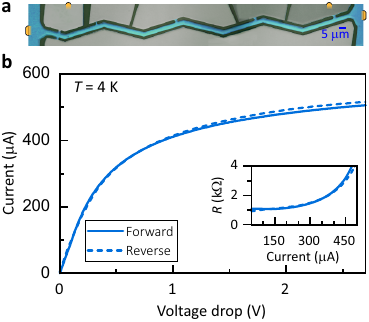}
    \caption{\textbf{No rectification in a reference sample without loops.} \textbf{a,} False-color optical image of the reference device. \textbf{b,} \textit{I-V} characteristics of the device demonstrating no rectification. Inset shows the resistance of the reference sample as function of DC current.}
    \label{fig:4}
\end{figure}

%No rectification in straight channels
Finally, to isolate the features of electron transport in Tesla valves, we fabricated the reference devices with $W\approx2$~$\mu$m that reproduce the geometry of the main channel in Tesla valves but contain no loops [Fig.~\ref{fig:4}].
As expected, their \textit{I-V} characteristics are symmetric and similar to that of Tesla valve under forward bias [Fig.~\ref{fig:1}d]. This confirms that the observed rectification arises solely from the Tesla valve geometry, whose looped design promotes head-on \textit{e-e} collisions leading to large reverse resistance.

% Outlook

Together, our findings reveal that the hydrodynamic description transcends classical fluids: a rectification mechanism reported in fluidic Tesla valves \cite{nguyen2021} operates equally in its electronic analogue. The asymmetric geometry places the created devices in the class of geometric diodes, yet the electron Tesla valve demonstrates remarkably high performance, with a diode efficiency $Di$ up to $40$, which is several times greater than the typical $Di\lesssim2$ reported for other geometric rectifiers \cite{nguyen2021a, wang2024}. Their planar structure yields low intrinsic capacitance, a natural advantage for high-frequency operation \cite{han2025}. These results open a path towards exploring electron Tesla valves at terahertz frequencies, where their performance could be benchmarked against existing high-speed rectifiers \cite{auton2016, auton2017}.

\section*{Acknowledgments} The study was funded by the Russian Science Foundation (Grant No. 22-12-00343-$\Pi$).

\section*{Authors contributions} D.I.S. and D.A.P. suggested, carried the project and analyzed the experimental data. D.A.P. and A.G.P. supervised the project. D.I.S. performed measurements with the technical support of E.Yu.Z. A.A.S. and A.K.B. fabricated samples. D.I.S. and D.A.P. wrote the manuscript. D.A.P., A.G.P. and A.A.S. contributed to the discussion.

\section*{Methods}
\subsection{Growth of heterostructure.} The GaAs/AlGaAs heterostructure, the basis of our samples, is grown in following steps: growth of Al$_{0.8}$Ga$_{0.2}$As layer ($400$~nm thick) on GaAs (001) substrate, growth of the heterostructure ($160$~nm thick) consisting of GaAs/AlAs superlattice, which is alternating layers of GaAs ($2.3$~nm thick) and AlAs ($1.15$~nm thick), and GaAs quantum well for electrons ($13$~nm thick). The structure was doped by Si $\delta$-layers with a density of $1.5\times10^{12}$~cm$^{-2}$ located at a distance of $40$~nm on both sides of the quantum well. The growth process is summarized in Extended Data Figure~\ref{ext_data:1}a.

\subsection{Lithography}

Large areas, including ohmic contacts, and the reference samples [Fig.~\ref{fig:4}] are patterned with photolithography and subsequent wet etching. Electron beam lithography was used to create the Tesla valves. To form the device boundaries, reactive ion etching is then performed through resist mask.

\subsection{Transport measurements}

Experimental samples are equpped with the Au/Ge/Ni ohmic contacts. Electrical measurements are performed in a dry VTI cryostat (Oxford Instruments Teslatron PT system) using 4-probe scheme, as shown in Fig.~\ref{fig:1}c, to exclude lead resistance. To measure Shubnikov-de Haas oscillations in Tesla valve [Extended~data~Figs.~\ref{ext_data:1}~b-d], we used lock-in technique at the frequency of $70$~Hz and the excitation current amplitude of $100$~nA.

\subsection{Characteristics of electron system: scattering lengths, electron temperature and Reynolds number.}

Momentum-relaxing length is extracted from the electron mobility measured in macroscopic Hall bar fabricated from the same GaAs/AlGaAs heterostructure as the Tesla valves. The Hall bar has a length and width of $L=50$~$\mu$m and $W=20$~$\mu$m, respectively [Extended~data~Fig.~\ref{ext_data:1}e]. The mobility $\mu$ is converted from the Hall bar resistance $R$ as $\mu=1/(neR)\times L/W$. Both values are shown in Extended~data~Figs.~\ref{ext_data:1}~f,g. Then, the momentum-relaxing length $l_\mathrm{MR}$ is obtained from the mobility as $l_\mathrm{MR}=mv_F\mu/e$ ($m$, $v_F$ are effective electron mass in GaAs and Fermi velocity, respectively).

Electron-electron scattering length, shown in [Fig.~\ref{fig:2}d], is calculated using the well-known formula of Giuliani and Quinn \cite{giuliani1982}:
$$l_{ee} = \frac{4\pi}{k_F} \left( \frac{E_F}{k_B T} \right)^2 \left[ \ln \left( \frac{E_F}{k_B T} \right) + \ln \left( \frac{2q_{TF}}{k_F} \right) + 1 \right]^{-1}.$$
Here, $E_F=\pi\hbar^2n/m=24$~meV is Fermi energy, $k_F=\sqrt{2mE_F}/\hbar\approx2.07\times10^8$~m$^{-1}$ is Fermi wavenumber, $q_{TF}=2me^2/(\hbar^2\varepsilon_\mathrm{GaAs})\approx2\times10^8$~m$^{-1}$ is inverse screening length for a 2DEG. This expression is confirmed \cite{sarypov2025b, egorov2024} to accurately describe \textit{e-e} scattering in studied 2DEG in GaAs quantum wells.

To estimate the values of electron temperature $T_e$ and Reynolds number $Re$ in our samples, we use the known energy-loss rates for electrons in GaAs quantum well \cite{shah1985}. At $I\approx350$~$\mu$A, power input from electric field is $P\approx I\cdot V\approx 350$~$\mu$W [Fig.~\ref{fig:1}d] yielding energy-loss rate per particle of $P/(nS)\approx2\times10^{-10}$~W ($S\approx1\times100$~$\mu$m$^2$ is the estimated area of the valve), which corresponds to $T_e\approx 100$~K \cite{shah1985}. At this temperature, viscosity of electron liquid can be lower than $\nu\approx6\times10^{-2}$~m$^2$/s \cite{sarypov2025}, giving a Reynolds number
$$Re=\frac{v_dW}{\nu}\equiv\frac{I}{ne\nu}\gtrsim7.$$

\bibliography{mybib}

\setcounter{figure}{0}
\begin{figure*}[h]
    \centering
    \includegraphics{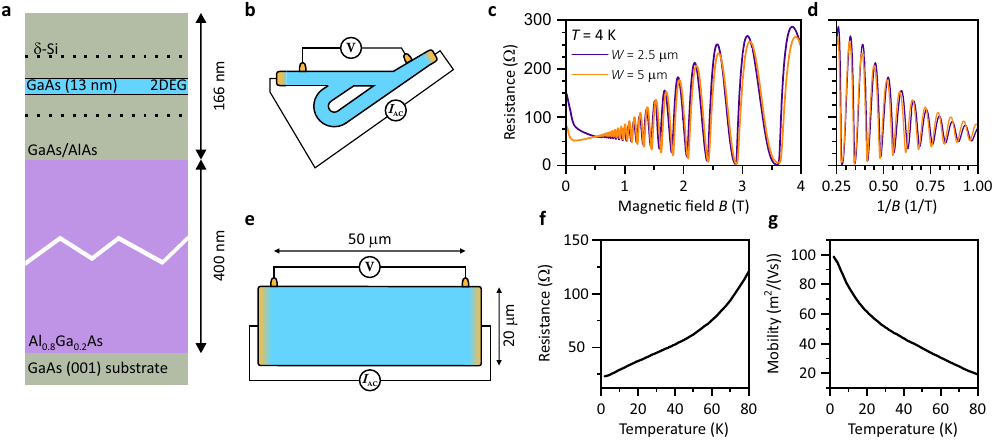}
    \renewcommand{\figurename}{Extended data Fig.}
    \caption{\textbf{Heterostructure and transport parameters.} \textbf{a,} Illustration of the GaAs/AlGaAs heterostructure. \textbf{b,} The scheme of the Tesla valve segment for magnetoresistance measurement. \textbf{c-d,} The resistance of the Tesla valve segment in magnetic field. \textbf{e,} The macroscopic Hall bar used to measure electron mobility. \textbf{f-g,} Temperature dependences of the Hall bar resistance and electron mobility.}
    \label{ext_data:1}
\end{figure*}

\end{document}